\documentclass[11pt]{article}
\usepackage{graphicx}
\usepackage{dcolumn}
\usepackage{bm}
\usepackage{comment}
\usepackage{color}
\usepackage[utf8]{inputenc}
\usepackage{amsmath}
\usepackage{jcappub}
\usepackage{bbm}
\begin{document}

\title{Transition Magnetic Moments and Collective Neutrino Oscillations:  Three-Flavor Effects and Detectability}

\author{Andr\'e de Gouv\^ea,}
\emailAdd{degouvea@northwestern.edu}
\affiliation{Department of Physics \& Astronomy, Northwestern University, Evanston IL 60208-3112, USA}
\author{Shashank Shalgar}
\emailAdd{shashank@northwestern.edu}


\note{NUHEP-TH/13-01}
\date{\today}

\abstract{
We demonstrate the non-negligible effect of transition magnetic moments on three-flavor collective oscillations of Majorana neutrinos in the core of type-II supernovae, within the single-angle approximation. We argue that data from a galactic supernova in conjunction with terrestrial experiments can potentially give us clues about the non-zero nature of neutrino transition magnetic moments if these are Majorana fermions, even if their values are as small as those predicted by the Standard Model augmented by nonzero neutrino Majorana masses.
}

\maketitle

\section{Introduction}

Over the past several years, the possibility of extracting neutrino properties from measurements of the flux of supernova neutrinos has been seriously considered. For example, collective neutrino oscillation effects, once properly taken into account, may allow one to, perhaps, robustly determine the neutrino mass hierarchy from measurements of the energy spectrum of electron-type neutrinos \cite{Dasgupta:2008my,Chakraborty:2008zp,Gava:2009pj,Fogli:2007bk,Duan:2006an,Raffelt:2007xt,Duan:2007fw,Duan:2007bt,Fogli:2008pt,Dasgupta:2009mg,Duan:2010bg}. It is known that there is a discontinuity in the electron neutrino spectrum for an inverted neutrino mass hierarchy induced by collective oscillations, usually referred to as a `spectral split'. 
An interesting feature of this phenomenon is its sensitivity to the mass hierarchy even for very small mixing angles. This can be understood in terms of a mean-field Hamiltonian, which is minimized for different configurations depending on the mass hierarchy of the neutrino \cite{Duan:2007mv, Raffelt:2007cb, Duan:2007fw, Duan:2007bt, Raffelt:2007xt, Fogli:2007bk}. In this language, the mixing angle serves to provide the initial perturbation necessary to take the system out of an unstable equilibrium configuration and the final result is hence almost independent from the value of the mixing angle, as long as it is nonzero. The prospect of being able to measure the mass hierarchy of the neutrinos irrespective of the value $\theta_{13}$ led to a lot of excitement, as it was identified as the only mechanism capable of such a measurement if it had turned out that $\sin^2\theta_{13}\lesssim 10^{-5}$. 
The recent discovery that $\theta_{13}$ is relatively large \cite{Abe:2011sj,Adamson:2011qu,Abe:2011fz,Ahn:2012nd} reveals that the hierarchy measurement is within reach of the current and next-generation of accelerator and atmospheric neutrino experiments (see, for example, \cite{Davies:2011vd, Blennow:2012gj}). 

We recently noted that a very small transition magnetic moment for the neutrinos can have a significant effect on collective neutrino oscillations in core-collapse supernovae \cite{deGouvea:2012hg} if these are Majorana fermions.\footnote{The issue of possible effects of transition magnetic moments in collective neutrino oscillations was raised previously in \cite{Dvornikov:2011dv}. The possibility that lepton-number violating effects could modify the dynamics of supernova explosions significantly was raised a long time ago, in \cite{Kolb:1981mc}.} 
Here we extend our analysis and consider the effects of transition magnetic moments in the context of three-flavor oscillations. Our motivation is threefold. It has been noted before in the literature that there are instabilities in collective neutrino oscillations that are artificial and a result of the two-flavor approximation \cite{Friedland:2010sc}. It is, therefore, important to make sure that the results reported in \cite{deGouvea:2012hg} are not an artifact of the two-flavor approximation. 
Furthermore, it is well-known that, after the collective oscillation regime, there is a non-trivial modification of neutrino flux spectra due to the standard MSW effect, which is also dependent on the neutrino mass ordering. This effect cannot be properly captured using the two flavor approximation, so in \cite{deGouvea:2012hg} we were unable to reliably compute the spectra of supernova neutrinos as these arrive on the surface of the Earth. 
Finally, the study of three-flavor collective oscillations, including Majorana neutrino magnetic moments contains more free parameters (e.g., there are three transition moments), a fact that may lead to qualitatively different phenomena.

We find that three-flavor effects do not ``wash out'' the results presented in \cite{deGouvea:2012hg}. We also compute expected supernova neutrino flavor spectra at the surface of the Earth, and estimate whether effects from nonzero transition magnetic moments can be ``seen'' in next-generation neutrino experiments. Finally, we also comment on the sensitivity of supernova neutrino oscillations to the Dirac $\mathcal{CP}$-violating phase in the lepton mixing matrix. It is known that the $\mathcal{CP}$-violating phase has no effect on the collective oscillations if the $\mu$ and $\tau$ neutrinos have the same initial fluxes \cite{Balantekin:2007es}. We find that this result no longer holds when the effects of transition magnetic moments and $\mathcal{CP}$-violating phases are considered simultaneously. Such effects, however, appear to be quantitatively too small to have phenomenological consequences in any realistic scenario. 

It is also important to make sure that the effects of transition magnetic moments persist in the more realistic multi-angle approximation \cite{Duan:2010bg,Sawyer:2008zs,Duan:2008eb,Duan:2008fd}, but here we restrict the analysis to the single-angle approximation. 
Past multi-angle studies, which do not take into account non-zero neutrino magnetic moments or other ``new physics''  reveal, however, that single angle calculations often provide excellent guidance regarding the importance of collective oscillations, including the order of magnitude of oscillation parameters above which we can expect a significant effect on the final flux. It seems reasonable to expect the same regarding non-zero neutrino transition magnetic moments. We highlight the fact that no multi-angle analysis of collective neutrino oscillations which included the effects of transition of magnetic moments has been performed to date. We are currently contemplating pursing this most challenging computation.  

We limit our analysis of collective effects to the later stages of supernova evolution ($\sim$~3.5 seconds after bounce), when the shock wave front is no longer interacting with the neutrinos, and ignore the time-dependency of the neutrino flux. Strictly speaking, our results apply only to a small fraction of the total neutrino flux, as most of the neutrinos in a core collapse supernova are emitted within the first second of the bounce. We focus on this late stage of evolution for simplicity and in order to properly highlight the effect of the neutrino magnetic moments. During earlier phases, very high matter densities render the calculation of the collective oscillations much more involved. The back-scattering of neutrinos due to ambient matter, for example, may play an important role in the collective oscillations and could modify the picture presented here qualitatively. As of the time of this writing, this topic remains unresolved even in the absence of transition magnetic moments \cite{Cherry:2012zw, Sarikas:2012vb}. We study the effect of transition magnetic moment on collective oscillation in the later stages of supernova evolution to avoid these and other complications. 

\section{Hamiltonian}

We use the same equations of motion as the ones used in  \cite{deGouvea:2012hg}, augmented to include three flavors of neutrinos and antineutrinos.

The vacuum Hamiltonian in the three-flavor case is a $6 \times 6$ matrix and a function of the neutrino mixing matrix $U$, the neutrino transition magnetic moments $\mu_{\alpha\beta}$, $\alpha,\beta=e,\mu,\tau$, the magnetic field inside the supernova and the mass eigenvalues -- a straight forward generalisation of the two flavor Hamiltonian considered in \cite{deGouvea:2012hg}:
\begin{eqnarray}
H_{vac} = 
\begin{pmatrix} 
H_{\theta} & H_{\mu B}\cr
-H_{\mu B} & H_{\theta}
\end{pmatrix},
\end{eqnarray}
where $H_{\theta}$ and $H_{\mu B}$ are given below in terms of the neutrino mass-squared differences, $\Delta m^{2}_{ij}$, $i,j=1,2,3$, the mixing angles that parameterize $U$ (we stick to the standard parameterisation \cite{Beringer:1900zz}), and the product of transition magnetic moments and the transverse magnetic field, $\mu B$.
\begin{eqnarray}
H_{\theta} &=& 
U
\begin{pmatrix}
0 & 0 & 0 \cr
0 & \Delta m_{12}^{2} & 0 \cr
0 & 0 & \Delta m_{13}^{2}
\end{pmatrix}
U^{\dagger},\\
H_{\mu B} &=& 
\begin{pmatrix}
0 & \mu_{e\mu} B & \mu_{e\tau} B \cr
-\mu_{e\mu} B & 0 &\mu_{\mu\tau} B  \cr
-\mu_{e\tau} B & -\mu_{\mu\tau} B & 0 \cr
\end{pmatrix}.
\label{H:mub}
\end{eqnarray}

Throughout, we use the following values for the mixing angles and the mass-squared differences \cite{Adamson:2011ig,Abe:2011sj,Adamson:2011qu,Abe:2011fz,An:2012eh}. We set all Majorana phases to zero, unless otherwise noted.
\begin{align}
\Delta m_{12}^2 &= 7.6\times 10^{-5}\ \textrm{eV}^2, &  \sin^{2}(\theta_{12}) &= 0.31, & \nonumber\\
\left|\Delta m_{13}^2\right| &= 2.5\times 10^{-3}\ \textrm{eV}^2, &  \sin^{2}(2\theta_{23}) &= 1, & \label{best:fit}\\
\delta &= 0^{\circ},180^{\circ}, &  \sin^{2}(2\theta_{13}) &= 0.1. & \nonumber
\end{align}
We will discuss the two different neutrino mass hierarchies (normal or inverted, characterized by the sign of $\Delta m_{13}^2$) individually. Although there are hints for the deviation of the atmospheric mixing angle from maximal (see \cite{Sousa:2011rw,Adamson:2011ku,Fogli:2012ua}), we only consider the maximal value of $\sin^22\theta_{23}$ in our simulation. Non-maximal $\theta_{23}$ values do not lead to qualitatively different results.

The matter effect due to ambient matter includes self-interactions, given by
\begin{equation}
H_{self} = \sqrt{2}G_{F}n_{\nu} \int dE ~ G^{\dagger}(\rho(E) - \rho(E)^{c*})G+\frac{1}{2}G^{\dagger}\mathrm{Tr}\left((\rho(E)-\rho(E)^{c*})G\right),
\label{h:self}
\end{equation}
where $G=\textrm{diag}(1,1,1,-1,-1,-1)$. The three-flavor matter effect due to ordinary matter (electrons) is $H_{mat}=\mathrm{diag}(\sqrt{2}G_{F}n_{e},0,0,-\sqrt{2}G_{F}n_{e},0,0)$. This self-interaction Hamiltonian is a straight forward generalisation of the one derived in \cite{deGouvea:2012hg}. A detailed discussion of the effect of the electromagnetic field on helicity changing amplitudes can also be found in \cite{Dvornikov:2011dv}.

\section{Neutrino fluxes and magnetic fields inside supernovae}

\begin{figure}[t]
\begin{center}
\includegraphics[width=7.75cm]{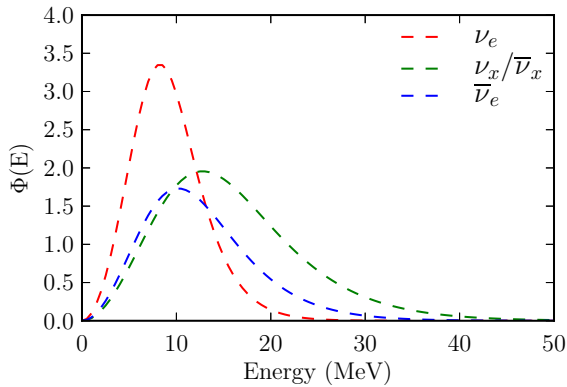}
\end{center}
\caption{Initial flux used in the simulation}
\end{figure}
We assume that the neutrino spectra have an initial distribution, at the neutrinosphere, that follows a Fermi-Dirac distribution with temperature ($T$), and chemical potential ($\eta$), dependent on the flavor $f=e,\mu,\tau,\bar{e},\bar{\mu},\bar{\tau}$ of the neutrino species. 
\begin{equation}
\phi_{f}(\mathbf{q}) \propto \frac{1}{\exp(q/T_{f}-\eta_{f})+1},
\end{equation}
where $\mathbf{q}$ is the neutrino momentum, which has magnitude $q$. We use chemical potentials, temperatures, and luminosities from the steep power law model described in  \cite{Keil:2002in}, with $p=10$ and $q=3.5$. As mentioned in the Introduction, this late-time flux, which we assume is time-independent, represents only a small fraction of total neutrino flux emitted in a core-collapse supernova event.

Integrating over the neutrino direction, we can express the flux as a function of $q$, which is, for all practical purposes, also the neutrino energy:
\begin{equation}
\phi_{f}(q) \propto \frac{q^{2}}{\exp(q/T_{f}-\eta_{f})+1}.
\end{equation}
The integral of $\phi_{f}(q)$ over $q$ yields the total number of neutrinos $\nu_f$ per unit volume. 

The effective neutrino number is dependent on the relative velocity between the neutrinos. Since the maximum relative angle between neutrino momenta fall as we move away from the neutrinosphere, the total number of neutrinos falls faster than inverse-distance squared. We normalize the total number of neutrinos at the surface of the neutrino sphere $R_{\nu}$, which we assume to be at a radius of 10 km. At the surface of the neutrino sphere, the effective luminosity is half that of the total luminosity. The effective neutrino number as a function of radius is
\begin{equation}
D(r) = \frac{1}{2}\left(1-\sqrt{1-\left(\frac{R_{\nu}}{r}\right)^{2}}\right).
\end{equation}
We normalize the total energy-luminosity as follows
\begin{equation}
n_{f}(r) = \frac{1}{2\pi R_{\nu}^{2}} \frac{L_{f}}{\langle E_{f} \rangle} D(r). 
\end{equation}
The density matrix is normalized to unity so that the coefficient $n_{\nu}$ in Eq.~(\ref{h:self}) is equal to the sum of $n_{f}$.
We are using $L_{e}=4.1\times 10^{51}$ ergs/sec, $L_{\bar{e}}=4.3\times 10^{51}$ ergs/sec and $L_{\mu,\bar{\mu},\tau,\bar{\tau}}=7.9\times 10^{51}$ ergs/sec \cite{Keil:2002in}. 

The ratio of the luminosity to the product of the transition magnetic moment and magnetic field, $B\mu$, determines whether the transition magnetic moment effects ``kick in'' or not. It is important to know the magnitude and direction of the magnetic field before conclusions about the value of the transition magnetic moment can drawn from the observations. We don't know of any astrophysical observations that can provide any clue regarding these quantities. However, we assume a reasonable profile for the magnetic field near the supernova core and, since only the transverse component of the magnetic field has any effect on the evolution of the supernova neutrino spectra, we assume it to be perpendicular to the line of sight of the supernova with respect to the Earth. 

The magnitude of the transition magnetic moments predicted in the Standard Model with massive Majorana neutrinos is around four orders of magnitude less than the would-be diagonal Dirac neutrino magnetic moments ($\mu_{D}\sim 3 \times 10^{-20} \mu_{B}$ for $m_{\nu} \sim 0.1$ eV), and calculable, as first discussed in \cite{Schechter:1981hw}, in terms of neutrino/charged-lepton masses and the parameters of the leptonic mixing matrix, including the Majorana phases. In more detail, given the values of the oscillation parameters listed in the previous section and using Eqs.~(83) and (84) from \cite{Broggini:2012df}, one can quickly compute the ``Standard Model'' expected ranges for the magnitude of the neutrino transition magnetic moments. For $m_{e},m_{\mu}\ll m_{\tau}$ and fixed $m_1=m_2=m_3=0.1$ eV, the transition magnetic moment components in the mass basis are, depending on the values of the Dirac and the Majorana phases,
\begin{equation}
|\mu_{ij}| = 
\begin{pmatrix}
0 & [ 0,3.1] & [0,3.3] \\
[0,3.1]  & 0 & [0,7.2] \\
[0,3.3] & [0,7.2]  & 0 \\
\end{pmatrix}\times 10^{-24}~\mu_{B}.
\end{equation}
We recall that the transition magnetic moment matrix for Majorana neutrinos is antisymmetric (see Eq.~(\ref{H:mub})), so the range of lower triangular components is uniquely determined by the range of upper triangular components. Furthermore, the various components of $\mu_{ij}$ are strongly correlated. The value of the transition moments in the flavor basis, $\mu_{\alpha\beta}$, $\alpha\beta=e,\mu,\tau$ (see Eq.~(\ref{H:mub})), are related to $\mu_{\ij}$: $\mu_{\alpha\beta}=U_{\alpha_i}^*U_{\beta j}^*\mu_{\ij}$.

It should be noted that, unlike previous studies of transition magnetic moments in the context of supernova explosions, for example in the context of explaining pulsar kicks \cite{Nardi:2000rr,Akhmedov:1997qb}, we use values of  magnetic fields and transition magnetic moments that are several orders of magnitude lower. Following the convention in \cite{deGouvea:2012hg}, we write the product of the transition magnetic moment and magnetic field in multiples of the product of the diagonal magnetic moment of a would-be Dirac neutrino and the magnetic field, denoted by $(\mu_{D} B)_{std}$, assuming a magnetic field that is inversely proportional to the square of the distance to the center of the explosion,
\begin{equation}
B(r)  = 10^{12} \left(\frac{50~\textrm{km}}{r}\right)^{2}~\textrm{gauss}.
\label{mag:field}
\end{equation}
Throughout the paper we assume the magnetic field profile described by Eq.~(\ref{mag:field}). The value of transition magnetic moment times magnetic field used in our simulations is of the same order of magnitude as predicted in the Standard Model for a neutrino mass of order 0.1 eV, $\mu B=10^{-4}(\mu_{D} B)_{std}$.

\section{Results}

For simplicity, we will consider the effects of  ``turning on'' one element of $\mu$ at a time, i.e., only one of $\mu_{e\mu}$, $\mu_{e\tau}$, and $\mu_{\mu\tau}$ is assumed to dominate the relevant phenomena at a given time. Hence we are ignoring the correlations between components of the transition magnetic moments that are, in the absence of new physics beyond Majorana neutrino masses, dependent on the neutrino masses and lepton mixing matrix, including Majorana phases. Hence, we have not considered the possible interference effects of two or all non-vanishing components simultaneously. We find that while the effects of the transition magnetic moment $\mu_{e\mu}$ is different from that of $\mu_{e\tau}$ or $\mu_{\mu\tau}$, these are qualitatively similar, and appear not be distinguishable in any conceivable realistic experiment. Hence, we only present results for a nonzero $\mu_{e\mu}$ and comment on what happens for a nonzero $\mu_{e\tau}$ or $\mu_{\mu\tau}$ when relevant. 

\begin{figure}[t]
\includegraphics[width=7.75cm]{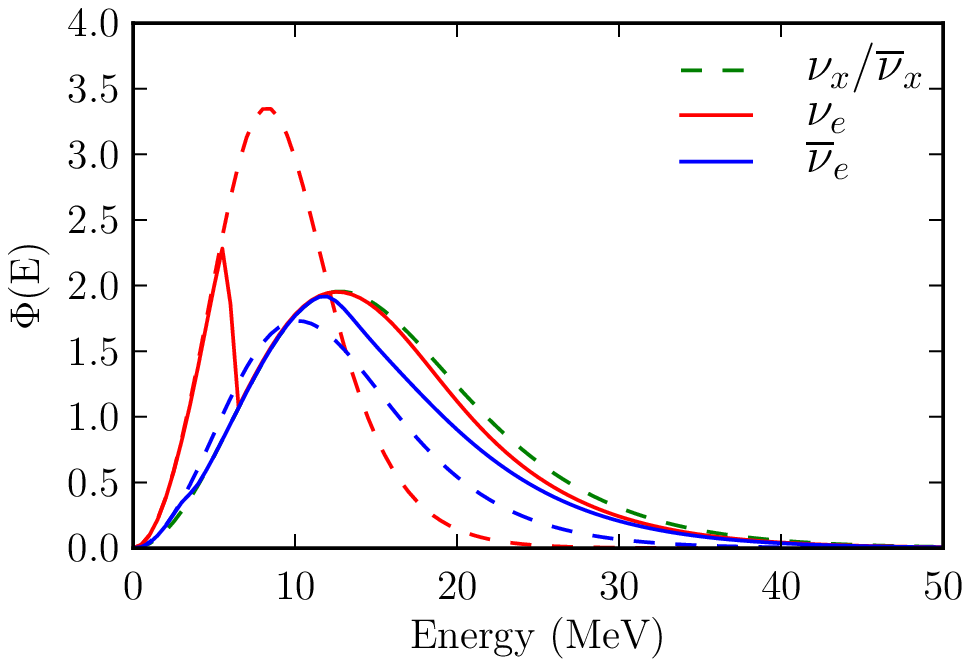}
\includegraphics[width=7.75cm]{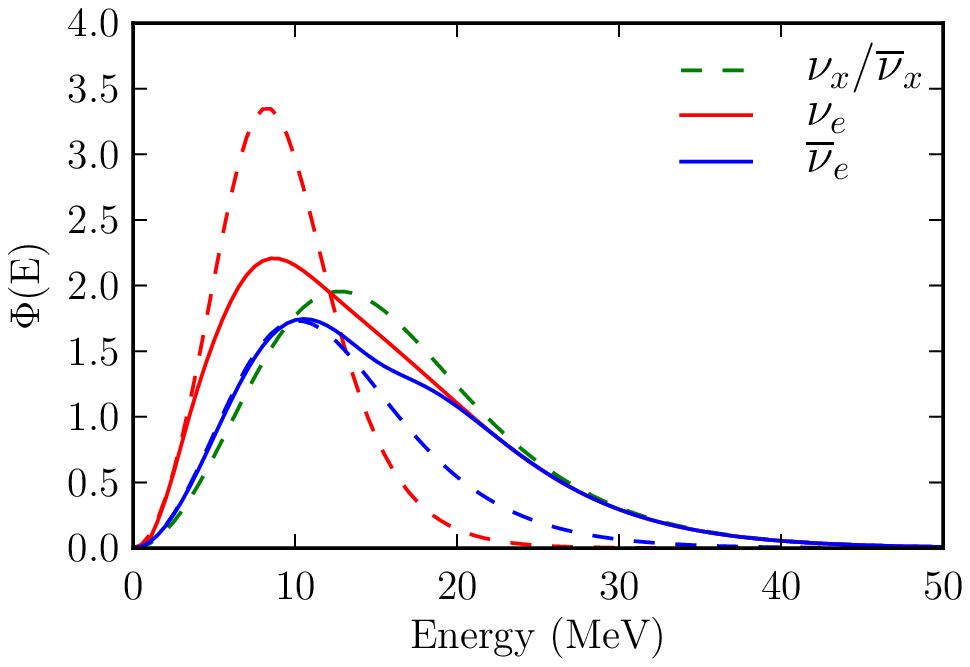}
\caption{Three-flavor spectra at $r=250$ km for an inverted (left) and normal (right) neutrino mass hierarchy, at best fit values of neutrino oscillation parameters(see Eq.~\ref{best:fit}) and no magnetic moment. The dashed (solid) lines represent initial (final) fluxes. Final $\nu_{x}/\bar{\nu}_{x}$, $x=\mu,\tau$  fluxes are not depicted. We use the same convention throughout the paper.}
\label{tf:both}
\end{figure}

\begin{figure}
\includegraphics[width=7.75cm]{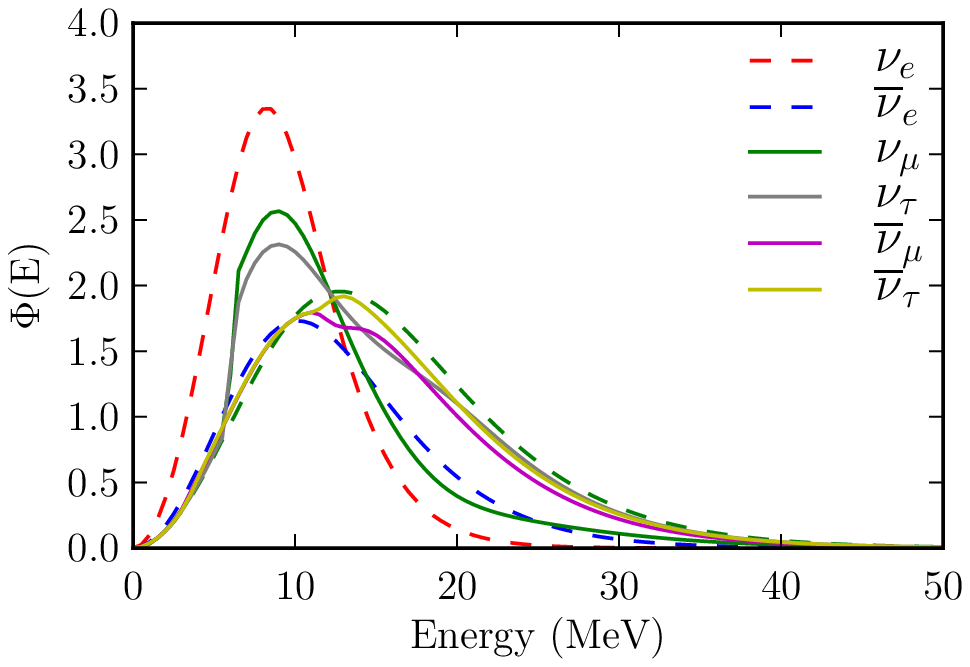}
\includegraphics[width=7.75cm]{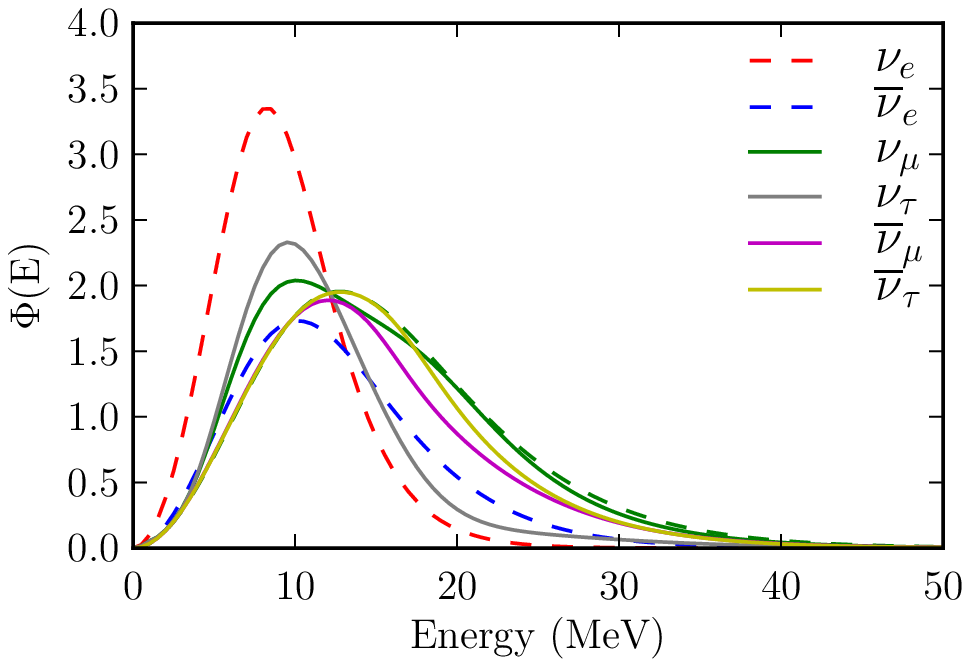}
\caption{Three-flavor spectra at $r=250$ km for an inverted (left) and normal (right) neutrino mass hierarchy, at best fit values of neutrino oscillation parameters and no magnetic moment. Final $\nu_{e}/\bar{\nu}_{e}$ fluxes are not depicted.}
\label{tf:bothx}
\end{figure}

\begin{figure}[!h]
\includegraphics[width=7.75cm]{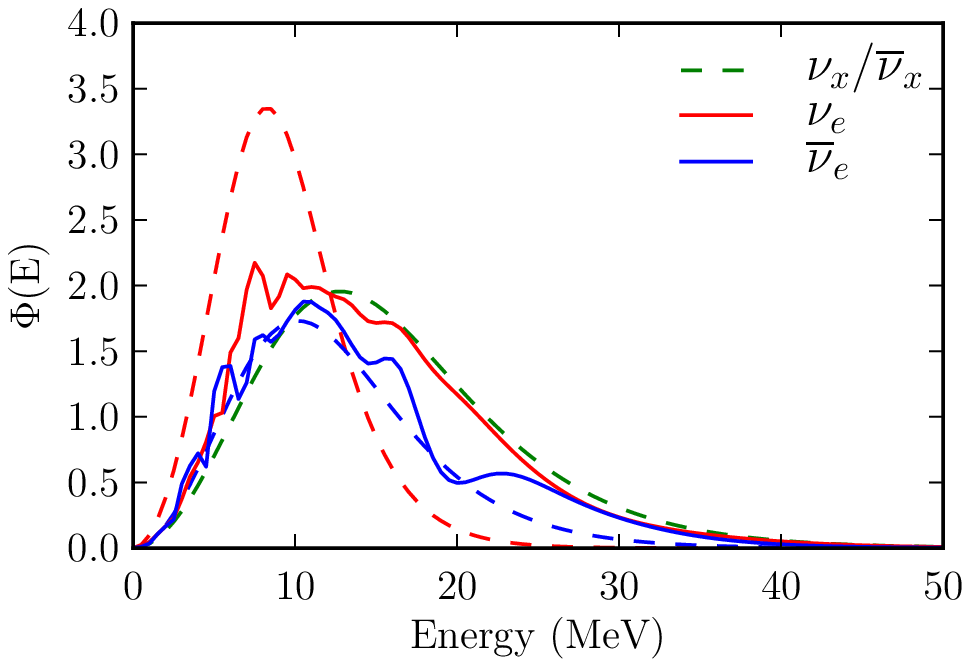}
\includegraphics[width=7.75cm]{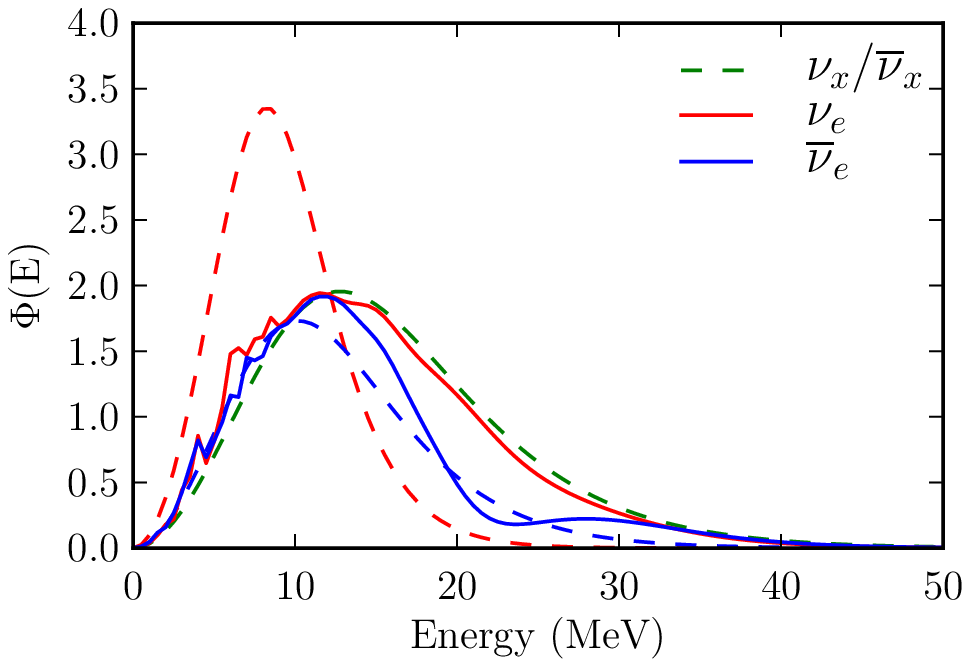}
\caption{Three flavor spectra at $r=250$ km for an inverted (left) and normal (right) neutrino mass hierarchy, at best fit values of neutrino oscillation parameters with $\delta=0^{\circ}$ and $\mu_{e\mu} B=10^{-4}(\mu_{D} B)_{std}$.}
\label{tfBemue4:del0}
\end{figure}

Figs.~\ref{tf:both}, \ref{tf:bothx} and \ref{tfBemue4:del0} depict the initial (dashed lines) and final (solid) neutrino fluxes as a function of energy and flavor, for a vanishing/non-vanishing $\mu_{e\mu}B$ and for a normal and inverted neutrino mass hierarchy (right and left, respectively). Here the final fluxes are computed at $r=250$~km, as in  \cite{deGouvea:2012hg}. Note that while Fig.~\ref{tf:bothx} depicts the $\nu_{\mu},\nu_{\tau},\bar{\nu}_{\mu},\bar{\nu}_{\tau}$ final fluxes, the other two, and all figures henceforth, depict only the $\nu_e$ and $\bar{\nu}_{e}$ fluxes, as these are the ones one is most likely to measure experimentally. Finally, for reference-purposes, all figures include the initial $\nu_e$, $\bar{\nu}_e$ and $\nu_x=\nu_{\mu},\nu_{\tau},\bar{\nu}_{\mu},\bar{\nu}_{\tau}$ fluxes. 

A comparison of  Figs.~\ref{tf:both} with Figs.~\ref{tfBemue4:del0} clearly reveals that the presence of even a tiny magnetic moment -- of order of the ones expected from Standard Model interactions -- leads to qualitatively different final spectra. As expected, the transition magnetic moments have a ``switch-on effect'' just like the elements of the neutrino mixing matrix, (e.g. $\theta_{13}$). We find that the $B\mu_{\mu\tau}$ switch-on values are higher than those for $B\mu_{e\mu}$ and $B\mu_{e\tau}$ but, for sufficiently large values of magnetic moments or neutrino luminosity, the final fluxes do not have features that qualitatively distinguish the various components of the transition magnetic moment matrix. Furthermore, for low enough values of the transition magnetic moment or neutrino luminosity, the final fluxes obtained with either nonzero $B\mu_{e\mu}$ or $B\mu_{e\tau}$ look exactly the same. Unlike the instability due to $\theta_{13}$ or the transition magnetic moment in the two-flavor approximation \cite{deGouvea:2012hg}, we do not see a clear `split' in the spectrum due to the transition magnetic moment for the three-flavor calculations; the three-flavor $\nu_e$ and $\bar{\nu}_e$ spectra have more structure. 

\begin{figure}
\includegraphics[width=7.75cm]{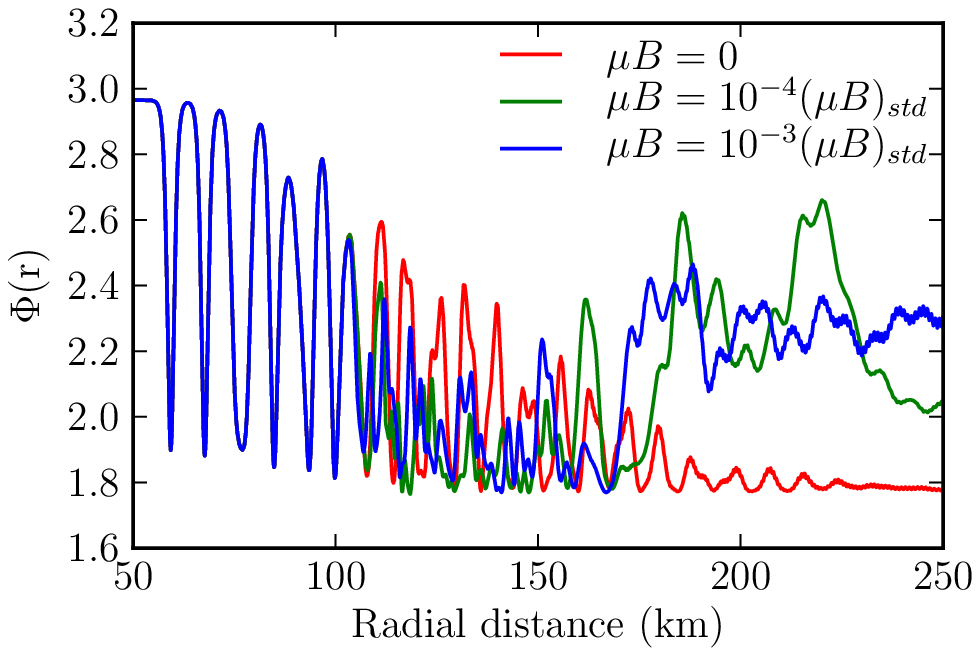}
\includegraphics[width=7.75cm]{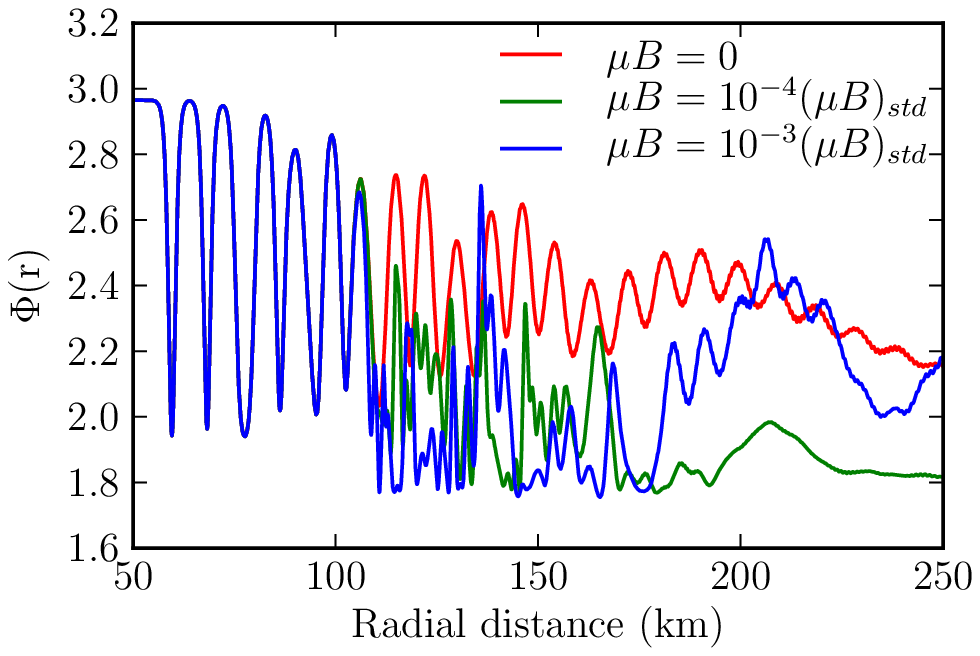}
\caption{$\nu_{e}$ flux as a function of distance for 10~MeV neutrinos and three values of $e\mu$ component of the neutrino transition magnetic moment -- $\mu_{e\mu} B=0, 10^{-4}(\mu_{D} B)_{std}, 10^{-3}(\mu_{D} B)_{std}$ -- for an inverted(left) and normal(right) neutrino mass hierarchy.}
\label{both:modes}
\end{figure}

Figure~\ref{both:modes} depicts the position-evolution of the $\nu_e$ flux for different values of $B\mu_{e\mu}$. In the case of a nonzero $B\mu_{e\mu}$, the oscillatory structure is a lot less regular, and the value of the flux is still changing quantitatively as a function of $r$ as $r$ approaches 250~km. Furthermore, at ``large'' $r$ values, the flux depends quantitatively on the magnitude of $B\mu_{e\mu}$, for the $B\mu_{e\mu}$ values in which we are interested.  

\begin{figure}[!h]
\includegraphics[width=7.75cm]{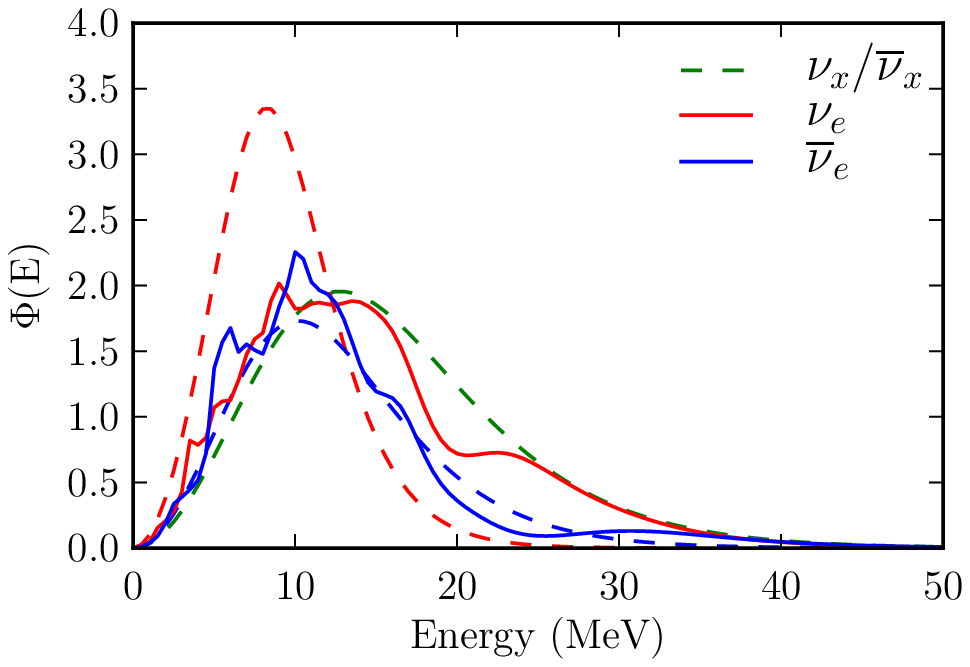}
\includegraphics[width=7.75cm]{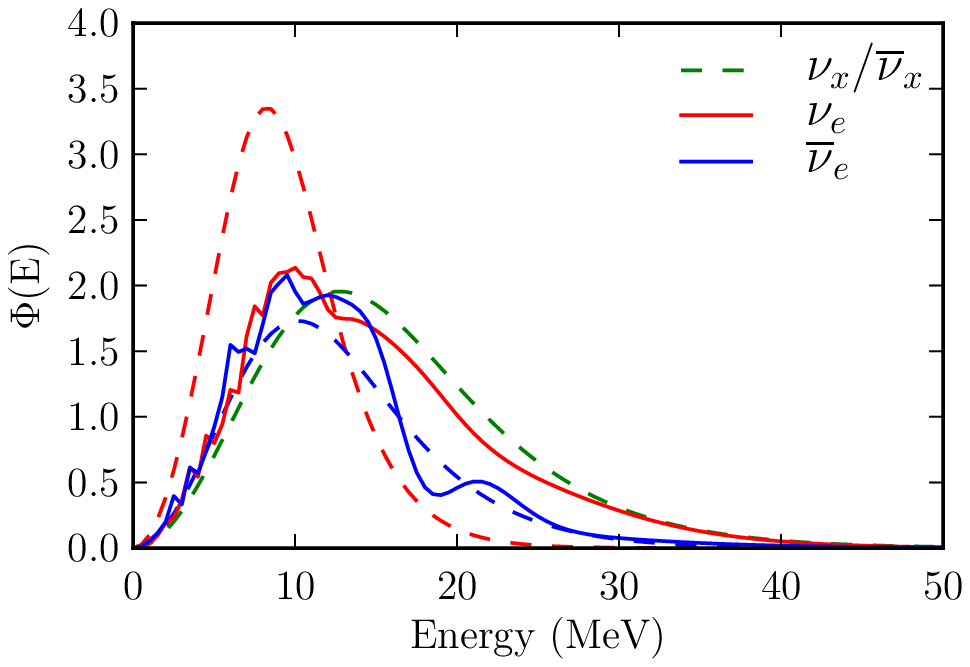}
\caption{Three flavor spectrum at $r=250$ km for an inverted (left) and normal (right) neutrino mass hierarchy at best fit values of neutrino oscillation parameters with $\delta=180^{\circ}$ and $\mu_{e\mu} B=10^{-4}(\mu_{D} B)_{std}$.}
\label{tfBemue4:del180}
\end{figure}

Fig.~\ref{tfBemue4:del180} is identical to Fig.~\ref{tfBemue4:del0} except for the value of the $\mathcal{CP}$-odd phase $\delta$, which is zero (Fig.~\ref{tfBemue4:del0}) or $180^\circ$ (Fig.~\ref{tfBemue4:del180}). It is easy to see that, quantitatively, the two sets of figures are not the same, which implies that, in principle, if the transition magnetic moment effects are not negligible, the flux of neutrinos from supernova should depend on $\delta$. 

In the absence of magnetic moment effects, since the survival probability of the electron type neutrino($P_{ee}$) is independent of the $\mathcal{CP}$ phase, the final electron-type neutrino flux ($\phi^{f}_{e}$), in the case of identical $\nu_{\mu}$ and $\nu_{\tau}$ initial fluxes($\phi^{i}$), is not affected by the $\mathcal{CP}$ phase \cite{Balantekin:2007es}:
\begin{eqnarray}
\phi^{f}_{e} &=& P_{ee}\phi^{i}_{e}+P_{\mu e} \phi^{i}_{\mu} + P_{\tau e} \phi^{i}_{\tau}, \nonumber\\
 &=& P_{ee}\phi^{i}_{e} + (P_{\mu e} + P_{\tau e})\phi^{i}_{\mu} \quad (\mathrm{for}\ \phi^{i}_{\mu}=\phi^{i}_{\tau}), \\
 &=& P_{ee}\phi^{i}_{e} + (1 - P_{ee})\phi^{i}_{\mu}. \nonumber
\label{flux:eqn}
\end{eqnarray}
A nonzero $\mu_{e\mu}$ or $\mu_{e\tau}$ component of the transition magnetic moment matrix, however, breaks the $\mu-\tau$ symmetry of the system, rendering the survival probability dependent on the $\mathcal{CP}$ phase. Furthermore, Majorana transition magnetic moments allow for ``neutrino--antineutrino'' transitions, such that
\begin{eqnarray}
\phi^{f}_{e} &=& \sum_{\alpha=e,\mu,\tau,\bar{e},\bar{\mu},\bar{\tau}} P_{\alpha e} \phi^{i}_{\alpha}.
\end{eqnarray}
The large lepton-number violating contribution from $P_{\bar{e}e}$, $P_{\bar{\mu}e}$, $P_{\bar{\tau}e}$ qualitatively changes the nature of the problem.

Comparing Figs.~\ref{tfBemue4:del0} and \ref{tfBemue4:del180}, it appears that a measurement of the $\mathcal{CP}$ phase via measurements of the supernova flux spectra would not be possible, considering the practical observational challenges such as the uncertainty in the initial flux spectra, limited statistics and limited energy resolution of the neutrino detectors. 

This effect of the Dirac $\mathcal{CP}$ phase is proportional to the total luminosity of the supernova and in the case of lower energy-luminosity the effect vanishes. The reason for this is that, for low-enough neutrino luminosities, the system becomes, effectively, a two-flavor system, and no $\mathcal{CP}$-violating effects can be accommodated. While it appears inconceivable to use this dependence on the $\mathcal{CP}$ phase as a method of measuring the $\mathcal{CP}$ phase, this result underscores the importance of measuring the $\mathcal{CP}$ phase in terrestrial experiments in order to extract as much physics as possible from supernova neutrino fluxes. The non-trivial effect of the $\mathcal{CP}$ phase is also a good example of why it is imperative to search for supernova-model-independent measurable quantities in the context of collective oscillations.

\begin{figure}[!h]
\includegraphics[width=7.75cm]{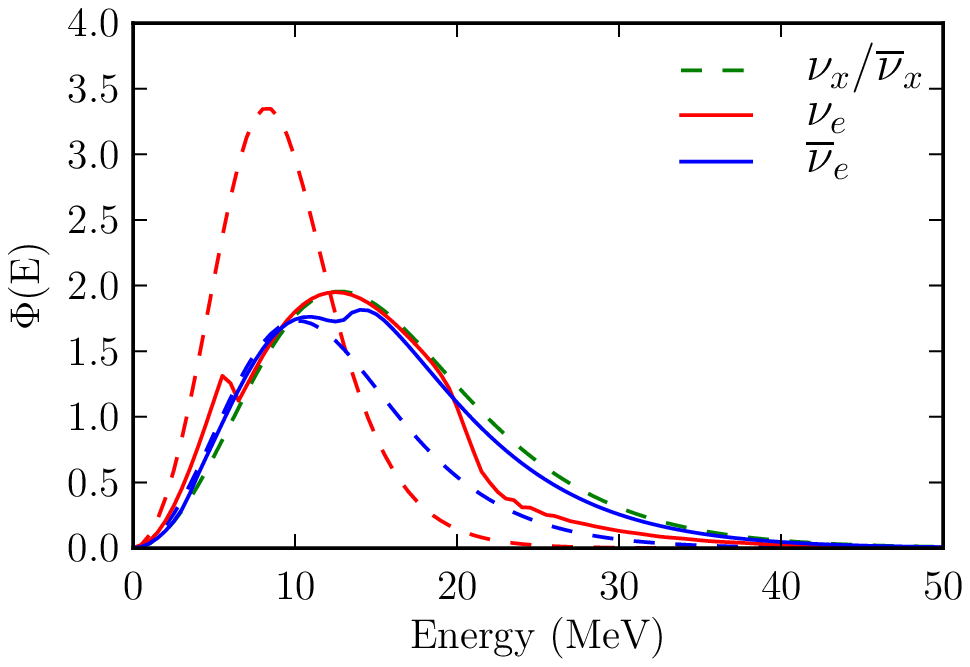}
\includegraphics[width=7.75cm]{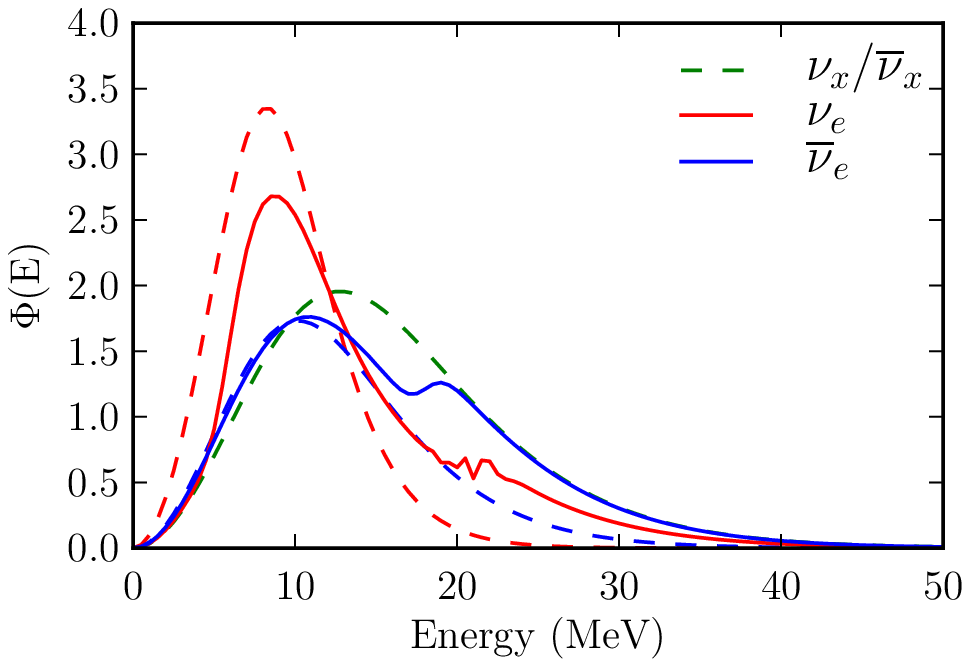}
\caption{Three flavor spectrum averaged over $r=24000-25000$ km for an inverted (left) and normal (right) neutrino mass hierarchy, at best fit values of neutrino oscillation parameters, $\delta=0^{\circ}$ and $\mu_{e\mu} B=0$.}
\label{tf:full}
\end{figure}

\begin{figure}[!h]
\includegraphics[width=7.75cm]{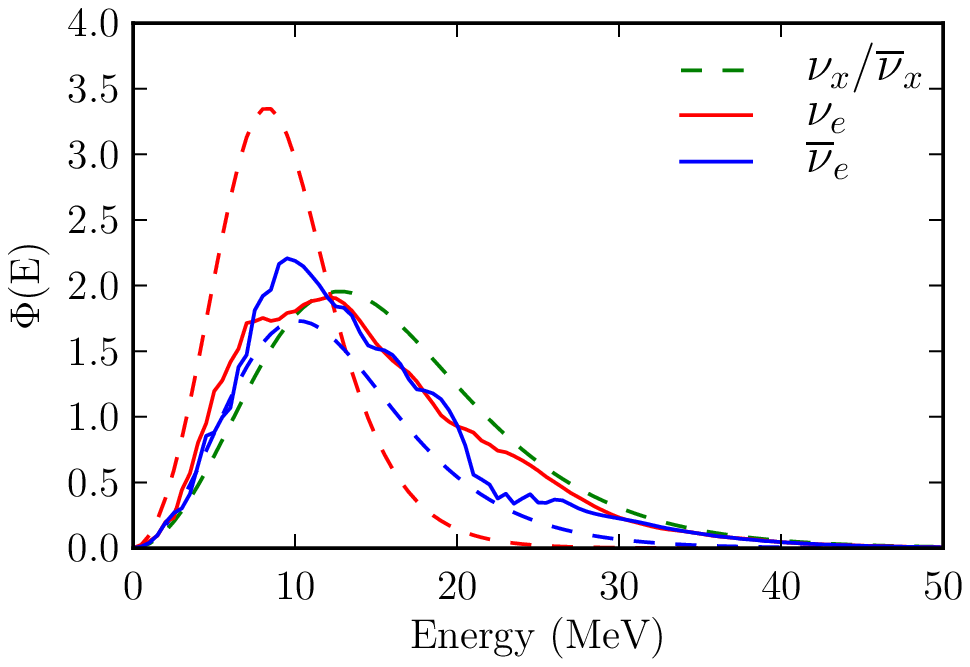}
\includegraphics[width=7.75cm]{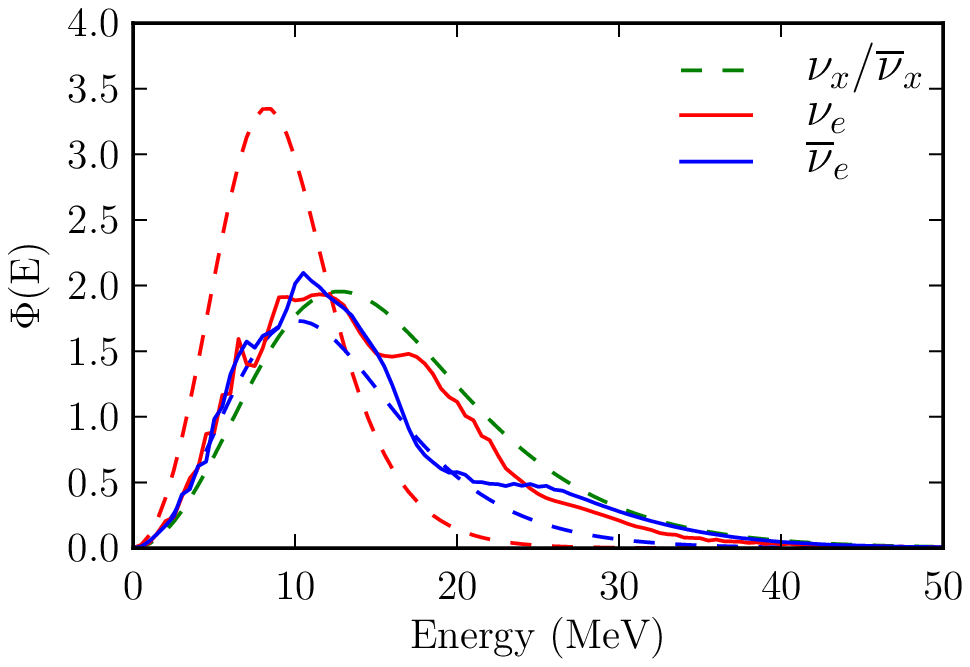}
\caption{Three flavor spectrum averaged over $r=24000-25000$ km for an inverted (left) and normal (right) neutrino mass hierarchy, at best fit values of neutrino oscillation parameters, $\delta=0^{\circ}$ and $\mu_{e\mu} B=10^{-4}(\mu_{D} B)_{std}$.}
\label{tfBemue4:full}
\end{figure}

In order to compute fluxes that are representative of the spectra we would observe in a detector on/in the Earth,  it is necessary to study the evolution of the neutrinos in the post-collective regime. In order to find the neutrino spectra beyond the standard MSW region, we calculate the average neutrino fluxes for $r$ values between $24000$ and $25000$ km. These fluxes are depicted in Figs.~\ref{tf:full} and \ref{tfBemue4:full}, for a zero and nonzero $B\mu_{e\mu}$, respectively. We average over a finite length in order to get rid of rapid oscillations and thus take into account the effect of kinematic decoherence that occurs when the neutrino travels a very long distance. Apart from the MSW effect in the supernova at relatively larger radii, there are also potentially significant matter effects in the Earth (see, e.g., \cite{Lunardini:2000sw,Lunardini:2001pb,Takahashi:2001dc,Dighe:2003vm,Mirizzi:2006xx,Dasgupta:2008my}). Here we ignore Earth matter effects. These are dependent on the position of the supernova in the sky with respect to the neutrino detector. 

After the MSW effect is taken into account, when magnetic moment effects are negligible, the prominent split present in case of an inverted neutrino mass hierarchy becomes less dramatic (see Fig.~\ref{tf:full}). This is consistent with comparable results discussed in \cite{Dasgupta:2007ws}. In the presence of 
a nonzero transition magnetic moment, standard matter effects, while significant, do not qualitatively modify the effects due to the magnetic moment (see Fig.~\ref{tfBemue4:full}). 

Quantitatively, all results are very sensitive to the initial neutrino flavor spectra. Considering the uncertainty in these initial neutrino spectra, the most promising way of experimentally searching for effects of the transition magnetic moments would be to compare the flux of electron neutrinos to that of electron antineutrinos. Here we consider the ratio of electron neutrino to electron antineutrino events that would be observed in an Earth-bound experiment. Since the cross-section increases with energy, this ratio is dominated by the flux in the high energy region. In the absence of the effect of a transition magnetic moment, there is an excess of antineutrino flux over neutrino flux (see Fig.~\ref{tf:full}). This is not the case when the magnetic moment effect is present (see Fig.~\ref{tfBemue4:full}).  We find that this is a robust feature in all comparable simulations that include the effects of a transition magnetic moment. 

We estimate the ratio of events by integrating the product of the charged-current cross-section in argon and the final fluxes over the energy. The charge current interaction of electron neutrinos and anti-electron neutrinos results in the formation of unstable nuclei, 
\begin{eqnarray}
\nu_{e} + \phantom{i}^{40}\textrm{Ar} & \rightarrow & \phantom{i}^{40}\textrm{K}^{*} + e^{-},  \\
\bar{\nu}_{e} + \phantom{i}^{40}\textrm{Ar} & \rightarrow & \phantom{i}^{40}\textrm{Cl}^{*} + e^{+},
\end{eqnarray}
 the gamma decays of which could be distinguished by virtue of their different $Q$ values and transition states, thus allowing charge identification \cite{Cocco:2004ac}. We use the cross-sections depicted in Fig.~1 of \cite{Cocco:2004ac}. 
 
 For a liquid argon-type detector, the ratio of electron neutrino to electron antineutrino events $R_{e/\bar{e}}$ , assuming the final fluxes depicted Fig.~\ref{tf:full}, for inverted and normal hierarchy is $R_{e/\bar{e}}=22.60$ and $R_{e/\bar{e}}=20.05$, respectively. Assuming the final fluxes depicted in Fig.~\ref{tfBemue4:full}, the same ratio is $R_{e/\bar{e}}=35.91$ and $28.67$, respectively. Depending on the mass hierarchy, and whether it is known, a measurement of $R_{e/\bar{e}}$ for late-arriving neutrinos\footnote{Keep in mind that, strictly speaking, our results apply to supernova neutrinos that are produced a few seconds after bounce.} at around the 10\% level\footnote{The uncertainty should include  all ``systematic uncertainties", including those related to the modeling of the neutrino source.} would allow one to conclude that the data are not compatible with zero neutrino transition magnetic moments, in the absence of other new physics.
 
 The ratio described is, arguably, a very crude observable one can construct out of future observations of supernova neutrinos, but we can, nonetheless,  still possibly extract evidence for a non-zero transition magnetic moment in this fashion. That all ratios are much greater than one is an artifact of the fact that liquid argon has a much larger neutrino-nucleon cross-section than antineutrino-nucleon cross-section. As already mentioned, the increase in the cross-section with energy gives a greater weight to the high energy region of the flux spectra. Robust features associated to magnetic moment effects in this high energy region provide the most reliable information when it comes to distinguishing the flux with and without transition magnetic moment. The robustness of these features, along with a more detailed study of potentially observable effects of nonzero magnetic moments, deserves further investigation with different models for the initial fluxes, etc, beyond the intents of this paper.

On the flip side, we would like to point out that a nonzero magnetic moment effect may hinder our ability to determine neutrino oscillation parameters using (even very precise) measurements of the neutrino fluxes from supernova explosions. For example, in the case of a  normal mass hierarchy and ignoring the effect of the transition magnetic moment (Fig.~\ref{tf:full}(right)) there is a significant excess of electron neutrinos over antineutrinos in the 10 MeV region. However, once the effect of a transition magnetic moment is included (Fig.~\ref{tfBemue4:full}), this distinguishing feature disappears. In isolation, the absence of an excess of electron neutrinos at around 10~MeV does not serve as evidence for an inverted neutrino mass hierarchy. It may simply be a consequence of a nonzero magnetic moment effect. Note that even if the initial flux spectra were known very well, it appears very challenging to tell Fig.~\ref{tfBemue4:full}(right) from Fig.~\ref{tfBemue4:full}(left). Keeping in mind that magnetic moment effects may be absent even if the neutrino magnetic moments are not, since these depend on the properties of the magnetic field inside the supernova, it is clear that the extraction of information on neutrino oscillating parameters from supernova explosions is a very nontrivial endeavor.\footnote{We remind readers that our results are not meant to address observables related to the time-dependency of the supernova neutrino flux, or the flux of neutrinos that are produced within one or two seconds of the ``bounce.''} 


Before concluding, we would like to highlight that, as is well-known, large enough values of the neutrino transition magnetic moments, combined with  the standard MSW effect inside astrophysical objects, can significantly modify the neutrino oscillation patterns. This has been studied quite extensively in the context of the solar neutrino problem (See for Eg.~\cite{Akhmedov:1993fv, Akhmedov:1996ec, Balantekin:1990jg}) as well as resonant spin-flavor oscillations in core-collapse supernovae for large magnetic moments \cite{Ando:2003pj,Lim:1987tk,Akhmedov:1992ea}. The same effect can also play a role inside supernovae in the post collective oscillation regime. Here, however, we are interested in values of the magnetic moment that are so small that all potential  such effects are negligible. We also note that one of the most stringent limits on the magnetic moment of a Dirac neutrino, for example, comes from supernova SN1987A (see, e.g., \cite{Barbieri:1988nh,Ayala:1998qz,Kuznetsov:2009zm}). The induced active--sterile neutrino transitions make it  easier for the supernovae to loose energy in the form of sterile neutrinos. Here, of course, we assume the neutrinos are Majorana fermions, and such processes are absent.


\section{Conclusion}

We have computed the effects of nonzero Majorana neutrino transition magnetic moments on the oscillation of supernova neutrinos. Expanding on the results presented in \cite{deGouvea:2012hg}, we have included the existence of the three confirmed neutrino species and included all known information regarding neutrino oscillation parameters. In this three-flavor scenario, we were allowed to not only compute the neutrino oscillation patterns at small radii, where collective effects dominate, but also include the standard MSW effects that occur at larger distances from the center of the supernova. This way, we were able to compute, for a fixed set of initial neutrino spectra, the different neutrino flux spectra, as a function of energy and flavor, when the supernova neutrinos arrive at the Earth's surface. These are depicted in Figs.~\ref{tf:full} and \ref{tfBemue4:full}. We also discussed a very coarse measure of whether Magnetic moment effects can be seen by Earth-bound neutrino detectors. 

We confirm that the qualitative results obtained in the two-flavor approximation persist when the existence of all three neutrino species is acknowledged. More quantitatively, we find that transition magnetic moment values of the order of those predicted by the Standard Model augmented by Majorana neutrino masses can have significant effects on collective neutrino oscillations. This result depends on the neutrino luminosity and the magnitude and direction of the magnetic fields inside the supernova explosion. We highlight, however, that the values assumed here are in the ``middle of the pack'' as far as a scan of supernova explosion models is concerned. To the best of our knowledge, the oscillation of supernova neutrinos appears to be the only physical phenomenon where standard-model-sized Majorana neutrino magnetic moments may lead to an observable effect.

We presented detailed results assuming that the $e\mu$-component of the neutrino magnetic moment tensor is nonzero. We find that the effect of the other components, when turned on individually, is qualitatively similar. We also find that the Dirac $\mathcal{CP}$-violating phase $\delta$, once magnetic moment effects are taken into account, has an effect on the evolution of the supernova neutrino fluxes, unlike the case of zero magnetic moments. However a visual examination of Figs.~\ref{tfBemue4:del0} and \ref{tfBemue4:del180} suggests that the effects are too subtle to be of any importance from a measurement point of view.

Our results provide yet another ``barrier'' as far as the enterprise of measuring the unknown neutrino oscillation parameters -- especially the neutrino mass hierarchy -- via the measurement of supernova neutrinos is concerned. In a nutshell, the flux of supernova neutrinos depends very sensitively on not only the neutrino oscillation parameters but also on quantitative details of the supernova explosion (initial neutrino luminosities and temperatures, etc) and other neutrino properties. The subject of this paper, transition magnetic moments, is a very non-exotic -- massive Majorana neutrinos have nonzero transition magnetic moments -- example of the latter. Other subtle but non-exotic effects exist, including, for example, relative $\mu-\tau$ refraction terms. These can also significantly affect collective neutrino oscillations in supernovae \cite{EstebanPretel:2007yq} and have been ignored here. While disentangling all the different effects appears to be a daunting task, it is important to keep in mind that the potential pay-off is huge. 

If it turns out that it is not, in general and model-independently, possible to determine the neutrino mass hierarchy from precise measurements of the flux of supernova neutrinos, it is still true that such a measurement will provide invaluable information regarding supernova explosions and other neutrino properties. Indeed, it may be most productive to consider that the job of analyzing supernova neutrino spectra will be made easier, and the results more robust, once the neutrino oscillation parameters are well-measured using different neutrino sources, both natural (e.g. atmospheric neutrinos) and man-made (accelerator and reactor neutrinos). Indeed, if the mass-hierarchy were known, it would be significantly easier to robustly establish that neutrino transition magnetic moments are not zero by measuring the flux of electron-type neutrinos and antineutrinos. 

We would like to remind the reader that these conclusions are valid only for the collective oscillations in the late stages of the supernova evolution. The physics of collective oscillations in the early stages is a difficult endeavour. However, it is conceivable that we may be able to isolate the neutrinos from the later stages of evolution or gain more insight using the time variation of the neutrino flux as we continue to learn more about the physics of collective neutrino oscillations. While significant progress has been made, the phenomenology of collective neutrino oscillations is still in its infancy. 
Our goal here is to draw attention to fact that Majorana neutrino transition magnetic moment effects are not negligible, regardless of the existence of  ``new physics,'' and cannot be ignored in future phenomenological studies. 

The phenomenon of collective oscillations continues to surprise us because of its highly non-linear dependence on neutrino parameters. Although we have established the extraordinary sensitivity of collective oscillations to Majorana transition magnetic moments, several other issues remain to be explored. Multi-angle calculations with transition magnetic moments, for example, are necessary to get a more realistic picture of the effect of Majorana transition moments on supernova neutrino spectra at the detectors. Also, there needs to be a thorough investigation of these lepton number violating effects in collective oscillations on other possible supernova-related observables, such as the abundances of heavy elements via r-process nucleosynthesis.



\acknowledgments{
We would like to thank Georg Raffelt for many useful comments on the draft.
This work is sponsored in part by the DOE grant \#DE-FG02-91ER40684.
}
\bibliography{threef}
\bibliographystyle{JHEP}
\end{document}